\documentclass[letterpaper]{article}
\usepackage{arxive-modified}
\usepackage{times}
\usepackage{helvet}
\usepackage{courier}
\usepackage{url}
\usepackage{varioref}

\usepackage{multicol}
\usepackage{lipsum}

\usepackage[super]{nth}

\usepackage{subfig}
\usepackage{multirow}

\usepackage{amsmath, mathrsfs}
\usepackage{array} 
\usepackage{amssymb}

\usepackage{amsthm} 
\theoremstyle{plain} 		
\theoremstyle{definition} 	
\theoremstyle{remark}		

\usepackage{tikz} 

\usepackage[font=small,format=plain,labelfont=bf,textfont=it]{caption}

\usepackage{graphicx}
\setkeys{Gin}{width=\linewidth,totalheight=\textheight,keepaspectratio}
\graphicspath{{figures/}}
\usepackage{booktabs}
\usepackage{units}

\usepackage[ruled]{algorithm2e}

\DeclareMathOperator*{\argmax}{arg\,max}

\begin{document}

\title{How Do We Find Early Adopters Who Will Guide a Resource Constrained Network Towards a Desired Distribution of Behaviors?}

\author{Kaushik Sarkar \and Hari Sundaram\\
School of Computing, Informatics and Decision Systems Engineering\\
Arizona State University\\
Tempe, AZ 85281\\
}

\maketitle

\begin{abstract}
\begin{quote}
We identify influential early adopters that achieve a target behavior distribution for a resource constrained social network with multiple costly behaviors. This problem is important for applications ranging from collective behavior change to corporate viral marketing campaigns. In this paper, we propose a model of diffusion of multiple behaviors when individual participants have resource constraints. Individuals adopt the set of behaviors that maximize their utility subject to available resources. We show that the problem of influence maximization for multiple behaviors is NP-complete. Thus we propose heuristics, which are based on node degree and expected immediate adoption, to select early adopters. We evaluate the effectiveness under three metrics: unique number of participants, total number of active behaviors and network resource utilization. We also propose heuristics to distribute the behaviors amongst the early adopters to achieve a target distribution in the population. We test our approach on synthetic and real-world topologies with excellent results. Our heuristics produce 15-51\% increase in resource utilization over the na\"ive approach.
\end{quote}
\end{abstract}

\section{Introduction}
This paper investigates how costly, multiple behaviors spread in social networks where individual have resource constraints. These constraints could include time, money or any material resource relevant to adopting the behavior.  The problem is challenging because the problem of finding early adopters or seeds is known to be computationally intractable. We develop heuristics to find early adopters that maximize multiple behavior adoption over resource constrained social networks.

Many of the pressing challenges facing contemporary society concern sustainability and public health. For example, how can sustainable behaviors---such as reducing individual energy consumption---be encouraged? How can participation in activities that reduce overall healthcare costs---such as compliance with preventive care routines and leading healthy lifestyles---be supported? These questions are termed as \textit{collective action} problems in the social sciences~\cite{Ostrom1999}. 

We are motivated by collective action problems to answer questions such as: how does a person's limited resources, including time, money or lack of tangible resources like a car or a bicycle, affect how she participates in real-world activities? A person interested in adopting a behavior (e.g. taking newspapers to a recycling station or voting) may fail to do so, due to
lack of resources.  A person may be interested in adopting multiple behaviors,
but each behavior has a cost. Current models of behavior adoption lack the idea that individuals may have significant resource constraints that preclude them from successfully adopting behaviors in which they are interested. Resource constraints not only limit individual participation, but also shape how behaviors spread in a network. We are interested in maximizing the use of resources available in a social network towards adopting a set of behaviors.

In this paper, we develop a model of multiple behavior diffusion that captures the complex dynamics of multiple behavior adoption in resource constrained networks. Our work is the first of its kind, to the best of our knowledge to study the influence of individual resource constraints on multiple, costly behavior adoption. In our model, behaviors have associated costs and utilities that are independent of the individual. Mindful of the work by ~\cite{Aral09} and ~\cite{shalizi11}, individuals in our model evaluate a utility function for each behavior that combines intrinsic interest and social signals.  An individual adopts a behavior when she receives a social signal of sufficient strength, the behavior is of high utility and when she has the resources to do so. We use three metrics: unique number of participants, number of behaviors in the network and expected resource utilization. Then, we identify two problems: which seeds to select, and what behaviors they need to adopt to maximize each of the three metrics.

We develop several heuristics to identify seeds that maximize the expected resource utilization in the network. These heuristics are necessary since we show that the problem of seed selection to maximize expected utilization is NP-complete. We also develop a heuristic to address a simple question: how do we distribute behaviors amongst the seeds to achieve a target behavior distribution? We test our approach on synthetic and real-world topologies with excellent results. We show that two heuristics that evaluate Expected Immediate Adoption provide very good solutions to the seed selection problems. We show that setting the seed behavior distribution to be proportional to the target behavior distribution produces excellent results. Our heuristics produce 15-51\% increase in resource utilization over the na\"ive approach.

The rest of the paper is organized as follows. In the next section we review the relevant literature. In Section~\ref{sec:problem} we formally define our behavior diffusion model and the set of problems that we address in the paper. In Section~\ref{sec:heuristics} we define the main seed selections heuristics. In Section~\ref{sec:dist} we discuss different behavior distributions that we will test. In Section~\ref{sec:sim-exp} we describe the simulation experiments and the results. In Section~\ref{sec:disc} we discuss open issues and extensions.  Finally,  we present our conclusions in Section~\ref{sec:conclusion}.

\section{Related Work}
The literature on social diffusion processes in vast. Diffusion of innovation was studied in \cite{rogers62}. Subsequent work on modeling such processes through epidemiological models was carried out by \cite{bass69} which became well known as the \textit{Bass Diffusion Model} in the management sciences community. Another approach of explaining such processes through threshold based models were made popular by \cite{granovetter83}. In recent years the study of diffusion modeling has seen substantial development. \cite{Watts02} presents a threshold based model of global cascades and analyzes why certain networks may appear to be ''robust'' yet turn out to be fragile against such cascades.

Works of \cite{Domingos02} and \cite{Kempe03} initiated the study of the computational problem of seed selection in the context of a ''viral'' social diffusion process. \cite{Kempe03} formalized the algorithmic problem for \textit{Independent Cascade} and \textit{Linear Threshold} models, proved the intractability results and provided a greedy approximation algorithm to the problem. However the approximation algorithm incurred a huge computational cost in practice. \cite{Leskovec07} came up with CELF technique to reduce the simulation cost of the algorithm. \cite{Chen09} tried to address that problem by coming up with computationally cheap heuristics that match the performance of the approximation algorithm. A complementary data mining perspective of inferring the diffusion model parameters from the past interactions the was taken by \cite{Saito08}, \cite{Goyal10} and \cite{Mathioudakis11}. 

Our work is informed by these literature, but is markedly different in a number of aspects. Much of the existing literature is concerned with diffusion of a single influence, while simultaneous diffusion of multiple influences is a more realistic scenario. Moreover none of these works take constraint on user resources into account and thereby does not apply directly to our problem of long term adoption of behaviors. \cite{Bharathi07} and \cite{Carnes07} discuss the problem of multiple competing influences, but they also do not incorporate the resource constraints or the utility maximizing behavior of social agents into their models. To the best of our knowledge the present work is the first investigation of the seed selection problem for multiple behavior diffusion in a resource constrained social network.

Next we mention a few interesting critiques of the study of social diffusion. The study of social diffusion and information contagion has met with its fair share of criticism. \cite{Aral09} argues that in their observational study more than $50\%$ of the perceived behavior contagion can be attributed to homophily instead of social influence. However \cite{shalizi11} have shown that homophily and social influence are generically confounded in social diffusion processes and it in general not easy to distinguish between the two effects. We have tried to take this observation into account while developing our model where. In our model the diffusion process is not exclusively driven by the social influence effects but an individual's intrinsic characteristics (i.e. resource constraint) also plays an important part in the adoption decision making.  An important critique of the study of modeling mass adoption through epidemic like contagion models is put forward by \cite{Goel12}. They have analyzed a number of real world product diffusion events and found out that in most of the cases the diffusion stops within one degree of the initial adopting seed, thus drawing a sharp contrast with multi-step person to person contagion of influence. It is however unclear if their findings will generalize to collective action problems, where participating individuals express interest in adopting a behavior (e.g. eating healthy foods, take a flu shot). As we discuss in our open issues section (Section~\ref{sec:disc}), there may be confounding effects between the communication problem and the decision making problem. That is, behavioral diffusion can halt in the presence of poor communication.

\section{Problem Definition}\label{sec:problem}
In this section we introduce our behavior diffusion model and the computational problem that we will address. In the following sections first we shall present our model of multiple behavior diffusion in resource constrained networks. Then, we shall introduce metrics, including resource utilization, unique participation and behavior adoption to evaluate the behavior adoption.  Finally, we shall discuss the problem of selecting the early adopters, and the concomitant issue of distribution of behaviors across the adopters.

\subsection{A Model of Multiple Behavior Diffusion in a Resource Constrained Social Network} \label{sub-sec:model}
We now describe the model for each user, the properties of each behavior and the behavior adoption process. We represent the social network with an undirected graph $G=(V,E)$. Each node $v \in V$ of the graph $G$ represents an individual and an edge $e \in E$ between two nodes indicate a social relationship between the two individuals. 

We wish to spread $k$ behaviors in the social network. Each behavior $i$ has an associated cost $c^i$ and a utility $u^i$. The cost refers to the cost of adoption and the utility refers to the intrinsic utility gained by an individual by adopting this behavior. In a simplification, we assume that both the cost $c^i$ and the utility $u^i$ of behavior $i$ are intrinsic to the behavior and independent of the individual who adopts the behavior. Without loss of generality, we assume that $0 \leq c^i, u^i \leq 1$.

Individuals are resource constrained: an individual may have limited time, money or may not possess other material resources to adopt a behavior. Therefore, we assign a \textit{fixed} resource $r(v)$ for each individual $v \in V$ towards adopting behaviors. The resource satisfies $0 \leq r(v) \leq 1$. For example, if we assume that individuals' resources are independent and identically distributed then the resource value $r(v)$ can be assumed to be obtained from a uniformly distributed random variable $U(0,1)$.  Let $N(v)$ denote the set of neighbors of $v$ in the network. Then we assume that a neighboring node $u$ asserts a social influence on node $v$ with weight $1/|N(v)|$.  

An individual will adopt a behavior $i$ when she receives a strong social signal, has the resources to do so and when the behavior is of a sufficiently high utility.  An individual $v$ adopts a behavior $i$ when the social signal exceeds a threshold $\theta^i(v)$, where $0 \leq \theta^i(v) \leq 1$. We assume that each individual $v$ has a different, fixed, threshold for each behavior, and that each threshold is obtained from a uniformly distributed random variable $U(0,1)$.  An individual $v$ can adopt a behavior $i$ provided the cost $c^i$ is less than $r(v)$, the available resources. The payoff $p^i(v)$ for a behavior $i$ is defined as the weighted sum of the intrinsic utility $u^i$ and the local network utility $l^i(v)$. That is, $p^i(v) = wu^i + (1 - w)l^i(v)$. Where, $w$ denotes the relative weight of the intrinsic utility and where $l^i(v)$ denotes the sum of influence weights---the social signal---exerted on $v$ by its neighbors who have adopted behavior $i$. If there are multiple behaviors that can be adopted, an individual will adopt a subset that maximizes payoff. 

Let us examine the diffusion of behavior over time, to illuminate the key ideas. The process takes place over discrete epochs\footnote{Notice that while actions in a network are asynchronous, we can choose an appropriate time granularity for analysis to assume synchronized decision making.}. We assume each node is aware of the behaviors adopted by her neighbors. The individual $v$ first identifies all candidate behaviors. A behavior $j$ is a candidate to be adopted if two conditions hold. First the social signal strength for behavior $j$ must exceed the threshold for that behavior at node $v$---$l^j(v) \ge \theta^j(v)$. Second, the individual $v$ must have the resources to adopt the behavior---$r(v) \ge c^j$.  The first condition is the familiar Linear Threshold (LT) model \cite{Kempe03}. Since there are multiple behaviors, the individual $v$ chooses a set of behaviors that will maximize the total payoff (i.e. the sum of payoffs $\sum_i p^i(v)$ over candidate behaviors) subject to the condition that the sum of the adoption costs of the behaviors are less than the resource constraint. That is $\sum_i c^i \leq r(v)$. At every epoch, the individual $v$ evaluates all behaviors, including behaviors already adopted, to evaluate payoff. 
The behavior diffusion process continues until no additional adoption is possible. 

In our diffusion model, we assume that the total resources available $r(v)$ at each node  are known, while the threshold for adoption $\theta$ for any behavior is unknown. This assumption is reasonable if when people are willing to make public their available resources to participate in a set of behaviors. This can arise say in a private, mobile social network app focused on adoption of healthy behaviors including wellness, healthy eating and exercise, where individuals join the network to participate in healthy behaviors but each individual is resource limited. An individual may declare that she has only one hour to spend on exercise each week, but would like to be nudged to participate in a health-related activity.

\subsection{Metrics}
We measure the effectiveness of the diffusion process with three metrics: total participation, total adoption and resource utilization. Since the behavior adoption is a stochastic process, we compute the expected value of each metric through simulation.

\subsubsection{Total Participation:}
This metric counts the the expected number of individuals who have adopted at least one behavior (i.e. become active) during the process. One goal for an advertiser who is interested in behavior adoption may be to maximize the total number of unique adoptees. Exact computation of this metric is an intractable problem (\#P-hard - \cite{Chen10}). 

\subsubsection{Total Adoption:}
In contrast to total participation, we need to keep track of the total number adoptions of any behavior during the diffusion process. This metric counts the expected number of adoptions over all the behaviors. Notice that since an individual can adopt more than one behavior, total adoption cannot be less than total participation. For the familiar single behavior adoption problem, these two metrics will have same value.

\subsubsection{Resource Utilization:}
This metric captures the \textit{efficiency} of the network to adopt costly behaviors. 
Not all resources available in a social network may be used for behavior adoption. This is  because individuals have variable resources, and they may be unable to adopt the subset of behaviors that fully takes advantage of their desire to participate because of two reasons. First, they may have many more resources than needed to adopt a behavior. Second, if their friends have limited resources, then the social signals that they receive will be about adopting low-cost resources, and hence a particular individual may never see costly behaviors in their social circle that they could potentially adopt.  let us assume that a node $v$ with resource $r(v)$ has adopted one or more behaviors. At the end of the diffusion process this individual has adopted a set of behaviors with total cost of adoption $s$, where  $s \le r(v)$. Therefore the individual has $r(v)-s$ amount of his resource remaining unused. \textit{Resource utilization} is the expected ratio of total utilized resource to the total amount of available resource of all the individuals in the social network. 

\subsection{The Seed Selection Problem}
There are two key problems: we need to identify the set of early adopters or seed nodes and we need to determine which behaviors ought to be adopted by each seed node. We assume that the number of initial adopters is small in comparison to the size of the network. This is reasonable as it corresponds to an advertiser with a finite budget to persuade the seeds to adopt. The behaviors adopted by the set of seed nodes have different implications on the metrics. If all nodes adopted the least costly behavior, for example, we would expect total participation to increase, but low resource utilization. The converse would be true in the case when seed nodes chosen such that all adopt the most expensive behavior. Next, we identify four subproblems; the first two refer to seed identification while the last two are concerned with behavior distribution in the seeds.

\subsubsection{P1: Resource Utilization Maximization:} Given a fixed seed budget $b$ and a fixed distribution of behaviors in the seed set, we want to select $b$ nodes in the network such that the resource utilization metric is maximized.

\subsubsection{P2: Total Participation (or Adoption) Maximization:} Given a fixed seed budget $b$ and a fixed distribution of behaviors in the seed set, we are interested in finding $b$ nodes in the network that maximize the total participation (or total adoption) in the network.

\subsubsection{P3: Determination of Optimum Behavior Distribution in the Seed Set:} Given a fixed seed budget $b$ and a lower bound on the number of adoptions of the lowest cost behavior $s_{min}$ (or a lower bound on the total participation $S_{min}$), what is the optimum distribution of behaviors in the seed set and the optimum set of $b$ seeds that will maximize the resource utilization while maintaining expected spread of $s_{min}$ for the lowest cost behavior (or $S_{min}$ for the total participation).


\subsubsection{P4: Obtaining Desired Distribution of Behaviors:} Given a fixed seed budget $b$, and a given fixed target distribution of behaviors $q$, how do we identify the $b$ seeds and the initial distribution of behaviors $p$ such that final distribution $p_T$ at time $T$ of behaviors in the population matches the target distribution $q$?

\section{Which Nodes Do We Pick?} \label{sec:heuristics}
In this section we develop heuristics to pick seed nodes to address each of the four seed selection problems. First we show that the seed selection problem is NP-complete. Then, we develop heuristics based on node degree and expected immediate adoption.
\subsection{NP-Completeness}\label{sec:NP}
It can be easily shown that the optimization problems P1 and P2 are NP-complete. We show that influence maximization problem for LT model, which is proven to be an NP-complete problem \cite{}, is a special case of P1. Let the number of behaviors $k=1$ and the cost of adoption of that behavior is also $1$. Each node $v$ is allocated resource $r(v)=1$. For these values of the parameters our multiple behavior diffusion model reduces to the LT model of influence propagation and resource utilization can be calculated as the ratio of the spread and total number of nodes in the network. So maximizing the resource utilization translates into maximizing the spread. Same transformation applies to problem P2 also since total participation (and total adoption) is identical to the spread in the one behavior case. Next, we propose a number of heuristics to solve the problem.

\subsection{Node Degree}
The social capital of an individual increases with increase in number of acquaintances. While the the nature of the connections and the specific structure of the network in which an individual is embedded matters, we can assume the node degree as a first order approximation to the ``influence'' of an individual. Hence heuristics based on node degree exploit this idea. We first discuss the basic heuristic and present some useful variants.

\subsubsection{Na\"ive:} In this variant we rank the nodes according to their degree and assign them different behaviors. This is a na\"ive extension of the high degree heuristic for the LT model \cite{Kempe03}. We test three variants of this heuristic. In the first variant, \textit{na\"ive degree with random tie breaking and no top up} (see Algorithm~\ref{algo:naive-rand-no-topup}) variant each seed node is assigned exactly one randomly chosen behavior only if its resource is sufficient for the cost of adoption of the behavior. In the second variant \textit{na\"ive degree with random tie breaking and top up}  each seed node is always assigned one randomly chosen behavior irrespective of its resource level. If its resource is not sufficient for adoption of the behavior we top up its resource so that it can bear the cost of adoption of the assigned behavior. In the third variant \textit{na\"ive degree with knapsack tie breaking}, each seed node is assigned all the behaviors that will maximize its utility subject to its resource constraint---each node will solve a knapsack problem to decide which set of behaviors to adopt. Notice that degree based heuristics are optimistic---it is possible that seed neighbors do not have resources to adopt the behavior of the seed.

\begin{algorithm}[t]
\SetAlgoNoLine
\KwIn{$G:=(V,E)$ - the social network, $\mathbf{b}$ - a vector of size $k$ containing number of required seeds for each of the behaviors}
\KwOut{$\mathbf{S}$ - a vector of size $k$ containing seed sets for each of the $k$ behaviors}
Let $V':=V$ and $\mathbf{S}:=\boldsymbol{\phi}$\;
\Repeat{$\mathbf{b}=\mathbf{0}$}{
    Select $v:=\argmax_u\{|N(u)| : u\in V'\}$\;
    $V':= V' - v$ \;
    Select $j$ uniformly at random from the set of behaviors $i$ that still need seeds to be assigned: $\{i:\mathbf{b}[i] \ne 0 \}$ \;
    \If{$r(v) \ge c^j$} {
       Set $\mathbf{S}[j] := \mathbf{S}[j] \cup \{v\}$ and $\mathbf{b}[j] := \mathbf{b}[j] - 1$\;
       Designate $v$ as an early adopter for behavior $j$\;
    }
   }
\caption{Na\"ive Degree Based with Random Tie breaking and No Top Up}
\label{algo:naive-rand-no-topup}
\end{algorithm}



\subsubsection{Neighbors With Sufficient Resource:} This heuristic takes in account both the degree and available resource of the neighbors when selecting the seed nodes. For each behavior $i$ we calculate $d^i(v)$ - the number of neighbors of a node $v$ with sufficient resource for adoption of $i$ (i.e. the number of neighbors $u$ with $r(u) \ge c^i$). Clearly $d^i(v)$ is a better indicator of the suitability of selecting $v$ as a seed for the $i$th behavior than just the node degree. In the \textit{degree and resource ranked} heuristic (see Algorithm~\ref{algo:degree-resource}) we compute $d^i(v)$ for all the nodes, rank them according to the value of this metric and select the required number of seeds for the $i$th behavior from the top of the ranking. If a node is selected as a candidate seed for more than one behaviors, we break the tie randomly and top up its resource so that it can adopt the randomly assigned behavior. We repeat the process until the required number of seeds are selected for all the behaviors.
Neither of the degree based heuristics provide any estimate of the effectiveness of the seed in terms of adoptions. We address this issue next.

\begin{algorithm}[t]
\SetAlgoNoLine
\KwIn{$G:=(V,E)$ - the social network, $\mathbf{b}$ - a vector of size $k$ containing number of required seeds for each of the behaviors}
\KwOut{$\mathbf{S}$ - a vector of size $k$ containing seed sets of required size for all the behaviors}
Let $d^i(v):=0$ for all $v\in V$ and $i \in \{1,\ldots,k\}$\;
\For{each $v \in V$}{
    \For{each behavior $i$}{
       \For{each neighbor $u$ of $v$}{
          \If{$r(u)\ge c^i$}{
             $d^i(v):=d^i(v)+1$\;
          }
       }
    } 
}
Let $V':=V$ and $\mathbf{S:=}\boldsymbol{\phi}$\;
\Repeat{$\mathbf{b}=\mathbf{0}$}{
    \For{each behavior $i$}{
       Let $T^i$ be the set of top $b[i]$ nodes from $V'$ in the decreasing sorted order of $d^i(v)$\; 
    }
    Let $T:=\cup_{i=1}^{k}T^i$\;
    Set $V':=V'\setminus T$ \;
    \For{each node $v$ in $T$}{
       Select $j$ uniformly at random from the set of behaviors $i$ with $v\in T^i$ \;
       \If{$r(v) \le c^j$} {
          Set $r(v):=c^j$\;
       }
       Set $S[j] := S[j] \cup \{v\}$ and $b[j] := b[j] - 1$\;
       Designate $v$ as an early adopter for behavior $j$\;
    }
   }
\caption{Degree and Resource Ranked Heuristic}
\label{algo:degree-resource}
\end{algorithm}

\subsection{Expected Immediate Adoption}
We compute the expected immediate adoption to estimate the influence of the seed. Notice that the exact computation of \textit{total} number of adoptions is \#P-hard for the LT model \cite{Chen10}, we can compute the one-step adoption in a straightforward manner.  Let $u$ be a neighbor of $v$. Hence the $v$ exerts a social influence of weight $1/|N(u)|$ on $u$. If $v$ is the only active seed in the network then the probability that $u$ will become active in the next time step is $1/|N(u)|$. This is because we have assumed that $u$'s threshold $\theta$ is assigned a value from a uniformly distributed random variable $U(0,1)$. Hence the expected number of adoptions after one time step, given that $v$ is the only active node at the beginning is $\scriptstyle1 + \sum_{u \in N(v)} \frac{1}{|N(u)|}$. For the multiple behavior case we will restrict the summation over those neighbors $u$ that have enough resource to adopt behavior $i$. We call this metric \textit{expected one step adoption of behavior $i$} and denote it by $e^i(v)$. Notice that two-step and three-step adoptions are difficult to analyze analytically---we would need to simulate the stochastic process to evaluate these two cases. The simulation will significantly increase the simulation cost In the next two sections, we describe two heuristics based on the expected immediate adoption metric. 
 
\subsubsection{Rank Based with Top-Up:} We rank all the nodes based on the value of $e^i(v)$ and choose the required number of seeds for behavior $i$ starting from the highest ranked nodes. We perform the same evaluation for all behaviors. If a node is selected as a candidate seed for more than one behaviors, then one of the behaviors is chosen randomly and assigned to the node. If the node does not have sufficient resource to adopt that behavior then its resource is topped up. The process continues until the required number of seeds are allocated to all the behaviors.

\begin{algorithm}[t]
\SetAlgoNoLine
\KwIn{$G:=(V,E)$ - the social network, $\mathbf{b}$ - a vector of size $k$ containing number of required seeds for each of the behaviors}
\KwOut{$\mathbf{S}$ - a vector of size $k$ containing seed sets of required size for all the behaviors}
Let $e^i(v):=1$ for all $v\in V$ and $i \in \{1,\ldots,k\}$\;
\For{each $v \in V$}{
    \For{each behavior $i$}{
       \For{each neighbor $u$ of $v$}{
          \If{$r(u)\ge c^i$}{
             $e^i(v):=e^i(v)+\frac{1}{|N(u)|}$\;
          }
       }
    } 
}
Let $V':=V$ and $\mathbf{S:=}\boldsymbol{\phi}$\;
\Repeat{$\mathbf{b}=\mathbf{0}$}{
    \For{each behavior $i$}{
       Let $T^i$ be the set of top $b[i]$ nodes from $V'$ in the decreasing sorted order of $e^i(v)$\; 
    }
    Let $T:=\cup_{i=1}^{k}T^i$\;
    Set $V':=V'\setminus T$ \;
    \For{each node $v$ in $T$}{
       Select $j$ uniformly at random from the set of behaviors $i$ with $v\in T^i$ \;
       \If{$r(v) \le c^j$} {
          Set $r(v):=c^j$\;
       }
       Set $S[j] := S[j] \cup \{v\}$ and $b[j] := b[j] - 1$\;
       Designate $v$ as an early adopter for behavior $j$\;
    }
   }
\caption{Expected Immediate Adoption Ranked}
\label{algo:ond-step-ranked}
\end{algorithm}

\subsubsection{Hill Climbing}
The hill climbing heuristic selects the seeds incrementally with the objective of maximizing the marginal increase of the one step spread. In this case while calculating the expected one step adoption of a node we do not consider the nodes that have already been selected as seeds. However, we account for their social influence when evaluating the expected immediate adoption.  As with the previous heuristic, if the node does not have sufficient resource then it is topped up so that it can adopt the assigned behavior.

\begin{algorithm}[t]
\SetAlgoNoLine
\KwIn{$G:=(V,E)$ - the social network, $\mathbf{b}$ - a vector of size $k$ containing number of required seeds for each of the behaviors}
\KwOut{$\mathbf{S}$ - a vector of size $k$ containing seed sets of required size for all the behaviors}
Let $V':=V$ and $\mathbf{S:=}\boldsymbol{\phi}$\;
\Repeat{$\mathbf{b}=\mathbf{0}$}{
    \For{each behavior $i$}{
       Let $T^i:=$Core-Hill-Climbing($i, b[i], S[i], V'$) \;
    }
    Let $T:=\cup_{i=1}^{k}T^i$\;
    Set $V':=V'\setminus T$ \;
    \For{each node $v$ in $T$}{
       Select $j$ uniformly at random from the set of behaviors $i$ with $v\in T^i$ \;
       \If{$r(v) \le c^j$} {
          Set $r(v):=c^j$\;
       }
       Set $S[j] := S[j] \cup \{v\}$ and $b[j] := b[j] - 1$\;
       Designate $v$ as an early adopter for behavior $j$\;
    }
   }
\caption{Expected Immediate Adoption Based Hill Climbing}
\label{algo:one-step-hill-climbing}
\end{algorithm}

\begin{algorithm}[t]
\SetAlgoNoLine
\KwIn{$i$ - the behavior, $b[i]$ - number of seeds required for the $i$th behavior, $S[i]$ - the set of already selected seeds for the $i$th behavior, $V'$ - the remaining population of nodes to choose new seeds from}
\KwOut{$T^i$ - the set of $b[i]$ newly selected seeds}
Let $e^i(v):=1$ for all $v\in V \setminus S[i]$ and $i \in \{1,\ldots,k\}$\;
\For{each $v \in V \setminus S[i]$}{
   \For{each neighbor $u$ of $v$ s.t. $u \in V\setminus S[i]$}{
      \If{$r(u)\ge c^i$}{
         $e^i(v):=e^i(v)+\frac{1}{|N(u)|}$\;
      }
   }
}
Let $T^i:=\phi$\;
\For{$j=1$ to $b[i]$}{
    Select $u:=\argmax_{v}\{e^i(v)|v\in V' \setminus T^i\}$
    $T^i:=T^i \cup \{u\}$\;
    \For{each neighbor $v$ of $u$ in $V' \setminus T^i$}{
       $e^i(v):=e^i(v) - \frac{1}{|N(u)|}$\;
    }
}
\caption{Core-Hill-Climbing}
\label{algo:Core-One-Step-Hill-Climbing}
\end{algorithm}

\section{How Do We Distribute the Behaviors?} \label{sec:dist}
It should be noted that in the case of multiple behavior diffusion metrics like resource utilization, total participation and total adoption depends not only on the choice of the seeds but also on the distribution of the different behaviors in the chosen seed set. We test following five different distributions of the behaviors in the seed set. In the \textit{highest cost behavior only} distribution we allocate all the seeds to the highest cost behavior and none to the other behaviors. In the \textit{proportional to cost} distribution the behaviors are distributed over the seeds in the ratio of their costs. \textit{Uniform} distribution divides the seeds equally amongst all the behaviors. In the \textit{Inversely proportional to cost} behavior distribution behaviors are distributed over the seeds in the inverse ratio of their costs. So the highest cost behavior gets the lowest number of seeds and the lowest cost behavior gets the highest number of seeds. Finally, in the \textit{lowest cost behavior only} distribution all the seeds are assigned to the lowest cost behavior and no seeds are given to the other behavior.  In problem P4 we are interested in achieving a target distribution. One heuristic is to assign the target distribution as the starting behavior distribution to over the seeds.

\section{Simulation Experiments} \label{sec:sim-exp}
In this section we describe different simulation experiments and compare the effectiveness of our proposed heuristics for seed selection and and for behavior distribution over seeds. We have implemented the multiple behavior diffusion model described in Section~\ref{sub-sec:model} and the heuristics discussed in Section~\ref{sec:heuristics} in the NetLogo Programming environment~\cite{Wilensky99}. In the following experiments we have assumed that we want to spread three behaviors ${b_1, b_2, b_3}$ with costs $c^1=0.2$, $c^2=0.5$ and $c^3=0.7$. We have assumed that behavior utility is proportional to cost. Hence our nominal utility values for the corresponding behaviors are $u^1=0.2$, $u^2=0.5$ and $u^3=0.7$. Finally, we assume that individuals' resources are independent and identically distributed i.e the resource $r(v)$ is uniformly distributed random variable $U(0,1)$ for all $v \in V$.

\subsection{Network Topologies}
We have used synthetic networks as well as a large real-world network for our experiments.  We synthesize network topologies through three social network generation models: preferential attachment \cite{barabasi99}; Small-world \cite{watts98} and spatially clustered \cite{Stonedahl08}. All the synthetic networks have $500$ nodes. In the preferential attachment network each new coming node adds one link to one of the existing nodes according to the in-degree distribution. The small world network formation starts with a regular circular lattice where each node is connected to next two nodes in the circular order. In the rewiring stage each edge is rewired with probability $p=0.2$. In the spatially clustered network average node degree is set to $10$. The three synthetic networks exhibit all the important properties---low effective diameter, power law degree distribution and high clustering---found in real world social networks. The real world data set is the ca-GrQc collaboration network form the SNAP network database \cite{Leskovec07a}. It is a collaboration network amongst authors who submitted their papers to the General Relativity and Quantum Cosmology category of e-print arXiv.org database. This network has $5242$ nodes and $28980$ edges.

The network types are abbreviated in the tables with experimental results as follows: PA (Preferential Attachment);  SW (Small World); SC (Spatially Clustered); COLL (the ca-GrQc collaboration network form the SNAP network database).
 
\subsection{Seed Selection}
In this section we compare the seven seed selection heuristics  described in Section~\ref{sec:heuristics} for different network topologies.  For the seed selection experiments, we fix the behavior distribution over the seeds: the behaviors are assumed to be uniformly distributed over the seeds. We use a specific fraction $\alpha$ of the population as seeds. In this experiment, we have used $\alpha=0.1$. This means that for synthetic networks, we use $b=51$ seeds\footnote{the number of seeds is a multiple of 3, since we have 3 test behaviors}, and $b=501$ for the real-world network.

The seven heuristics are abbreviated in the experimental results tables as follows: H1 (Random); H2 (Na\"ive Degree---No Top-up); H3 (Na\"ive Degree---Knapsack); H4 (Na\"ive Degree---Top-up); H5 (Degree and Resource Ranked), H6 (Expected Immediate Adoption---Ranked), H7 (Expected Immediate Adoption---Hill Climbing).

\begin{table}[htb]\footnotesize
\centering
    \caption{Maximum Possible Resource Utilization of different network types. Each node solves the knapsack problem and selects optimal behaviors. Then, we diffuse the behaviors. We are reporting the equilibrium values under two conditions: we fix the thresholds and vary topology (Network Average); we fix a random topology and vary thresholds (Threshold Average). Notice that the the quantum physics collaborative dataset, we cannot report a network average since the topology is fixed. }\label{tab:max-util}
    \begin{tabular}{ccc} \toprule
        Network & Threshold Average & Network Average \\ \midrule
        PA & 0.71 & 0.71		\\
        SW & 0.72 & 0.72 	\\
        SC & 0.73 & 0.73		\\
        COLL & 0.73 & N/A 	\\ \bottomrule
    \end{tabular}

\end{table}
Since seed selection sub-problems P2 and P3 are NP-complete (ref. Section~\ref{sec:heuristics}), determining the maximum possible utilization or total participation in the network for the given value of $b$ under uniform behavior distribution is computationally intractable. However, we can estimate the value of maximum possible utilization in the network if we assume that $b=N$, the case when each network node is a seed. First the nodes in the network adopt the subset of behaviors that maximizes their payoff subject to the resource constraint.  Then we let the diffusion process run till the network reaches equilibrium. The expected value of the resource utilization at this point will upper bound of resource utilization in that network and enables comparison with our heuristics. Table \ref{tab:max-util} provides the value of this maximum possible utilization for different networks. Notice that for three behaviors with costs $c^1=0.2$, $c^2=0.5$ and $c^3=0.7$, it is straightforward to show that the maximum utilization will be bounded by the value 0.78, assuming that the thresholds are obtained from $U(0,1)$. The fact that the simulation results are slightly lower that 0.78 is because nodes will ``align'' with their neighbors over time due to the social influence.

There are two sources of randomness in the synthetic network generation models:  behavior adoption thresholds at each individual for each behavior and network topology. Since each aspect is independent of the other, we have conducted two different types of simulations. In the first, we pick an arbitrary topology and vary individual thresholds over the different simulation runs. We term this as \textit{threshold average}. In the second type of simulation, we fix the individual thresholds, obtained from the uniform distribution, and vary the topologies over the simulations. We term this as \textit{network average}. Notice that the real-world dataset---ca-GrQc network---has a fixed topology and hence only one type of randomness: variation of the individual thresholds.  We use $5000$ independent runs of the diffusion process to obtain stable estimates for both threshold and network types of simulations.

\begin{table}[htb]\footnotesize
\centering
    \caption{Resource Utilization under Threshold / Network Average. Both versions of the Expected Immediate Adoption, heuristics H6, H7 give excellent results. The differences between the heuristics for the same type of average are statistically significant.}\label{tab:seed-selection-threshold}
    \begin{tabular}{ccccc} \toprule
        Heuristics & PA & SW & SC & COLL \\ \midrule
        H1 & 0.12 / 0.14 & 0.15 / 0.15  & 0.16 / 0.16 & 0.14 / -\\
        H2 & 0.22 / 0.24 & 0.16 / 0.17 & 0.16 / 0.17 & 0.18 / -\\
        H3 & 0.28 / 0.30 & 0.17 / 0.17 & 0.16 / 0.17 & 0.18 / -\\
        H4 & 0.32 / 0.33 & 0.17 / 0.17 & 0.17 / 0.18 & 0.19 / - \\
        H5 & 0.35 / 0.36 & 0.21 / 0.21 & 0.18 / 0.19 & 0.20 / - \\
        \textbf{H6} & 0.37 / 0.38 & 0.21 / 0.22 & 0.20 / 0.21& 0.28 / -\\
        \textbf{H7} & 0.37 / 0.38 & 0.22 / 0.22 & 0.21 / 0.22& 0.29 / -\\ \bottomrule
    \end{tabular}
\end{table}

Table \ref{tab:seed-selection-threshold} shows the estimated resource utilization of different networks for threshold and network average simulations for each of the seven seed selection heuristics. The two Expected Immediate Adoption heuristics (H6, H7) show the highest expected utilization. The differences between the heuristics (H6, H7) and the other heuristics are statically significant ($p < 0.01$) for the same type---threshold or network---of simulation. The table also reveals an expected result: the network average and the threshold averages are nearly identical for the same heuristic. In the next section, we discuss how to distribute behaviors over the seeds.

%

\subsection{Behavior Distribution over Seeds}

In this section we investigate the effects of the different behavior distribution heuristics across the initial seed set described in Section~\ref{sec:dist}. For this simulation, we use heuristic H7 (the one step spread based hill climbing heuristic) as it is the best performing seed selection heuristic. We designate the fraction of seeds to be early adopters to be $\alpha=0.1$. This means that we have $b=51$ for the synthetic networks and $b=501$ for the quantum physics collaboration network. As before, we compute the metrics under the threshold average and the network average simulations.

In the tables in this section, we shall use the following notation: Low (All seeds are assigned Lowest Cost Behavior); Inv. (the seeds are allocated behavior in Inverse proportion to behavior cost; Unif. (the behaviors are distributed Uniformly at random); Prop. (the behaviors are distributed Proportional to behavior cost); High ( all seeds are allocated the Highest cost behavior).

\begin{table}[htb]\footnotesize
\centering
\caption{Resource Utilization under Threshold / Network Average.  Among the behavior distribution heuristics, assigning every seed the lowest (highest) cost behavior results in the lowest (highest) utilization. Assigning seeds proportional to cost, works as well as the assigning everyone the highest cost behavior.  }\label{tab:behav-util-threshold}
    \begin{tabular}{ccccc} \toprule
        Heuristics & PA & SW & SC & COLL \\ \midrule
        Low & 0.23 / 0.23 & 0.14 / 0.13& 0.15 / 0.14 & 0.18 / -\\
        Inv.  & 0.33 / 0.35 & 0.20 / 0.21& 0.20 / 0.21& 0.27 / -\\
        Unif. & 0.37 / 0.38 & 0.22 / 0.22 & 0.21 / 0.22& 0.29 / -\\
        \textbf{Prop.} & 0.38  / 0.40 & 0.24 / 0.24 & 0.22 / 0.23& 0.31 / -\\
        \textbf{High} & 0.38 / 0.39 & 0.24 / 0.25 & 0.24 / 0.23& 0.31 / -\\ \bottomrule  
    \end{tabular}
\end{table}


\begin{table}[htb]\footnotesize
\centering
  \caption{Total Participation / Total adoption under Network Average for different behavior distributions over seeds. Seeds are chosen under heuristic H7. Notice that when all the seeds are the same behavior (Low, High), the number of unique participants and adoptions are identical. }\label{tab:behav-part-network}
    \begin{tabular}{ccccc} \toprule
        Dist. & PA & SW & SC \\ \midrule
       Low & 291.12 / 291.12& 166.26 / 166.26 & 178.78 / 178.78 \\
       Inv.  & 250.91 / 254.52& 146.36 / 150.09& 149.61 / 154.58\\
       \textbf{Unif.} & 234.00 / 236.46& 133.38 / 136.04& 132.27 / 135.77\\
       \textbf{Prop.} & 209.66 / 210.98 & 118.65 / 119.98& 113.29 / 114.91\\
       High & 144.49 / 144.49  & 93.79 / 93.79 & 86.01 / 86.01\\ \bottomrule        
    \end{tabular}
 \end{table}{\tiny }



Table~\ref{tab:behav-util-threshold} shows the resource utilization in different networks for the threshold average and the network average simulations. We have omitted the simulations for the threshold average case, due to space limitations.  Those simulations are qualitatively similar to Table~\ref{tab:behav-util-threshold}. We see that when each seed is either allocated the same low (high) behavior, the utilization is lowest (highest). This is unsurprising as we should expect high utilization to occur when we have high cost behaviors in the network. In Table~\ref{tab:behav-part-network}, we show the difference between the number of unique participants and the total number of behavior adoptions. Notice that when all the seeds have either the same low (or high) behaviors assigned to all of them, there is unsurprisingly no difference between the total number participants and the total number of unique adoptions. As Table~\ref{tab:behav-part-network} shows, change to the behavior distribution over the seeds alters the unique number of participants as well as the total adoption. Therefore the seed distributions need to be chosen with care, the appropriate metric in mind. Both uniform and propositional to cost behavior distribution methods seem to hit a sweet spot between utilization and behavior diversity.

\subsection{Achieving a Target Behavior Distribution}
In this section we present experiments on a specific heuristic to achieve a specific target distribution $q$. We propose the following heuristic: set the behavior distribution $p$ over the seeds, \textit{to be equal} to the target distribution $q$.

In this section we investigate the effect of seed set behavior distribution on the final distribution of behaviors in the population of active individuals. In addition to uniform, proportional to cost and inversely proportional to cost, 

We will test to if we can achieve the following six target distributions for the three behaviors in our simulations: $1:1:1$; $1:2:3$; $1:3:2$; $2:1:3$; $2:3:1$; $3:1:2$ and $3:2:1$. Notice that the target behavior distribution ratios are different from the behavior costs. As a reminder, the costs are $c^1=0.2$, $c^2=0.5$ and $c^3=0.7$.  We use the KL-divergence between target distribution and the actual behavior diffusion distribution to measure if our heuristic achieves the target distribution $q$.

\begin{table}[htb]\footnotesize
\centering
  \caption{KL Divergence, under threshold average conditions, between the equilibrium behavior diffusion and the target distribution $q$ when the initial behavior distribution over the seeds $p_{seeds}$ is set to the target distribution $q$, for the synthetic networks.}\label{tab:kl-div-threshold}
    \begin{tabular}{ccccc} \toprule
        Target Dist. $q$ & PA & SW & SC  \\ \midrule
        1:1:1 & 0.02 & 0.03 & 0.06 \\
        1:2:3 & 0.06 & 0.03 & 0.05 \\
        1:3:2 & 0.03 & 0.03 & 0.04 \\
        2:1:3 & 0.02 & 0.03 & 0.06 \\
        2:3:1 & 0.03 & 0.02 & 0.05 \\
        3:1:2 & 0.02 & 0.03 & 0.05 \\
        3:2:1 & 0.04 & 0.02 & 0.04 \\ \bottomrule 
    \end{tabular}
\end{table}

Table~\ref{tab:kl-div-threshold} shows that the heuristic of $p_{seeds} = q$ works well, and results in low KL-divergence between the equilibrium behavior distribution and the target $q$. One can improve on the heuristic by slightly underweighting the low-cost behaviors and slightly overweighting the high cost behaviors. This exploits the results of Table~\ref{tab:behav-part-network}, which shows that low-cost behaviors spread much farther than do high-cost behaviors.

\begin{figure}[htb]
\begin{centering}
\includegraphics[width=\columnwidth]{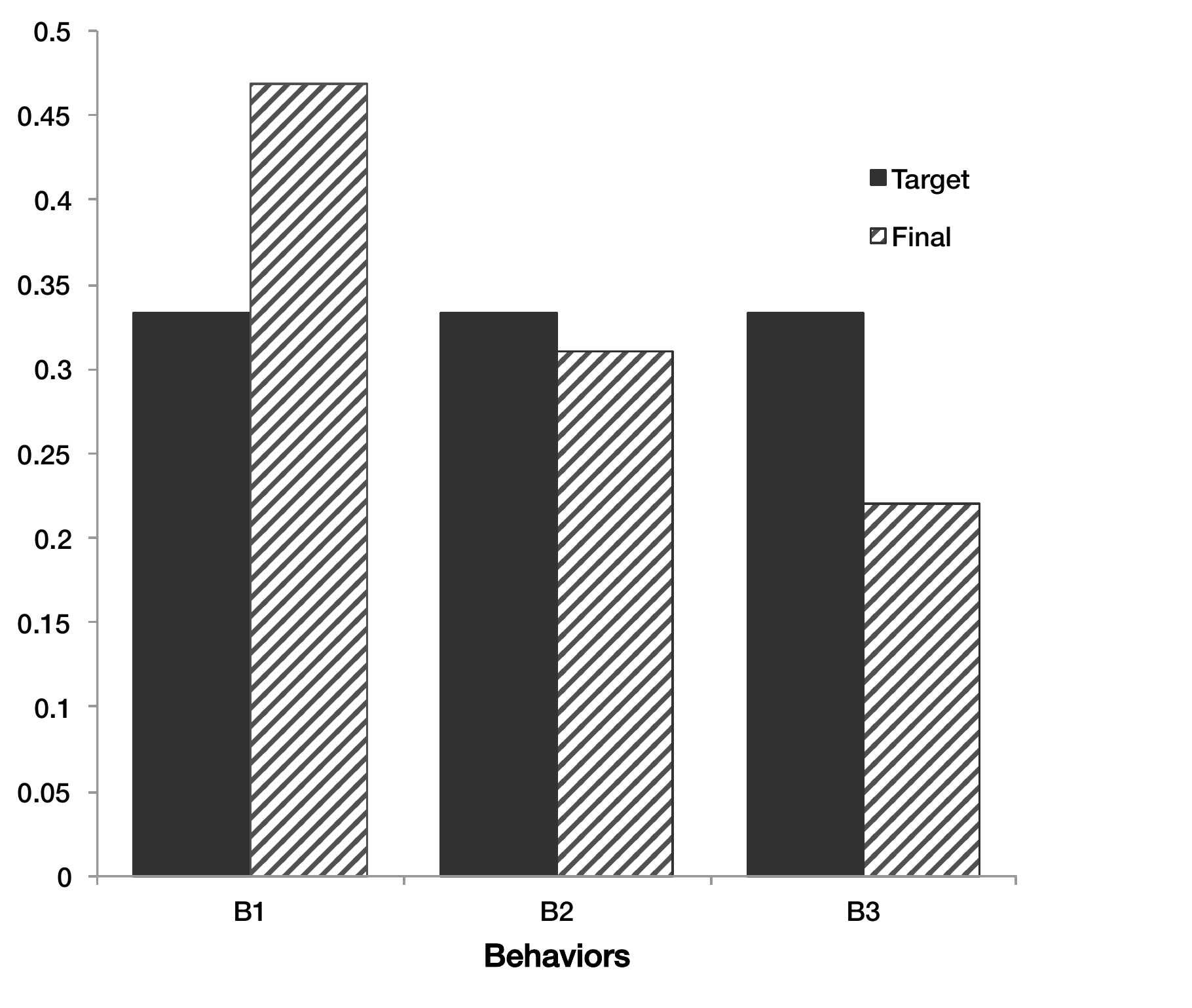}
\par\end{centering}
\caption{A comparison of the final equilibrium distribution and the desired target $q$ of uniform distribution of behaviors when the seed distribution $p_{seeds}$ is set to be uniform. The results are for the collaboration network (COLL). Low cost behaviors spread more than do high-cost behaviors. The KL divergence between the equilibrium behavior distribution and the target $q$ is 0.05.}\label{fig:snap-uniform}
\end{figure}

Figure \ref{fig:snap-uniform} shows the distributions for ca-GrQc collaboration network under threshold average type simulation with uniform initial distribution.  The KL divergence between the equilibrium behavior distribution and the target $q$ is 0.05. We can see that the lowest cost behavior increases its share and the highest cost distribution loses its share in the final distribution while the median cost behavior almost maintains its share.  Again the modified heuristic of underweighting low-cost behaviors, and overweighting high-cost behaviors will help decrease the KL divergence.

\begin{figure}[htb]
\includegraphics[viewport=1.5in 3.5in 6.75in 7.5in, width=\columnwidth]{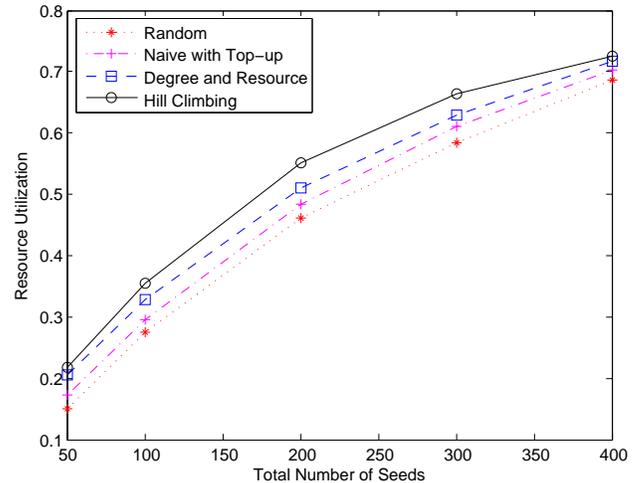}
\caption{The graph shows the comparison in resource utilization with the different number of seeds (that is, different values of the $\alpha$ parameter, for four heuristics. The heuristic H7, which maximizes expected immediate adoption performs the best. Notice that $l=50$ corresponds to $\alpha=0.1$.}\label{fig:comparison}
\end{figure}

Figure~\ref{fig:comparison} shows the comparison in resource utilization with different values of $\alpha$ the fraction of seeds in the network. We show four seed selection heuristics: random, na\"ive with resource top up, degree and resource ranked seeds and the expected immediate maximization heuristic with hill climbing (H7). In this simulation the behaviors were distributed uniformly across all the seeds. The heuristic H7 consistently outperforms the other three seed selection heuristics, for each value of $\alpha$.  

\section{Discussion and Open Issues}\label{sec:disc}
One of the main motivations of the present work was to develop a realistic model of the behavior diffusion process. There are many ways in which our work can be extended. Here we discuss about a few such possible extensions.

Our present model does not consider the role of behavioral inertia in the diffusion process. Often people are hesitant of adopting new behaviors because they cannot free their resources from practicing an old behavior which possibly has less value. This can be modeled in our framework by introducing an additional benefit for the already adopted behaviors. Another technique would be to introduce epidemic models such as SIRS to better model long-term behavioral adoption. 

In a network, we receive social signals from our friends, but there is noise because we miss messages and or we check them late. In modeling the behavior adoption problem, we have ignored the role of constraints in how they affect the production and consumption of messages from peers. Explicit consideration of the cost of social signaling would not only make the model more realistic and provide better bounds on the maximal resource utilization of the networks resources.

\section{Conclusions}\label{sec:conclusion}
In this paper we have considered the problem of seed selection to maximize resource utilization and to achieve a specified target distribution for multiple behavior diffusion processes.  We are motivated by collective action problems with applications to sustainability and public health. We have considered a social network where individuals are constrained by available resources for adoption of new behaviors. Our work is the first of its kind, to the best of our knowledge to study the influence of individual resource constraints on multiple, costly behavior adoption. Mindful of the confound between homophily and structural effects, individuals in our model respond to the social influence as well as the intrinsic utility of a spreading behavior. We have shown that the core optimization problems are NP-complete and provided novel heuristics for solving them. We have tested our heuristics against the random and na\"ive methods and have shown that our heuristics perform very well. We have also shown that assigning the target distribution as the behavior distribution over the seed set results in the final diffused behaviors being very close to the target distribution. Some of the open issues include the use of epidemic models for modeling long-term behavior adoption and incorporating the idea of noisy social signals in modeling behavior adoption.

\bibliography{refs}
\bibliographystyle{aaai}
\end{document}